\documentclass[12pt,preprint]{aastex}

\newcommand{\bdv}[1]{\mbox{\boldmath$#1$}}

\def\au{{\rm AU}} 
\def\kms{{\rm km}\,{\rm s}^{-1}}
\def\masyr{{\rm mas}\,{\rm yr}^{-1}}
\def\rel{{\rm rel}}
\def\geo{{\rm geo}}
\def\hel{{\rm hel}}
\def\fwhm{{\rm fwhm}}
\def\e{{\rm E}}
\def\bpi{{\bdv\pi}}
\def\bmu{{\bdv\mu}}

\def\muas{{\mu\rm as}}

\begin{document}
\title{{\it Kepler}-like Multi-Plexing for Mass Production of 
Microlens Parallaxes}

\author{
Andrew Gould\altaffilmark{1},
Keith Horne\altaffilmark{2}
}
\altaffiltext{1}{Department of Astronomy, Ohio State University,
140 W.\ 18th Ave., Columbus, OH 43210, USA; 
gould@astronomy.ohio-state.edu}
\altaffiltext{2}{SUPA, University of St Andrews, School of 
Physics \& Astronomy, North Haugh, St Andrews, KY16 9SS, UK;
kdh1@st-andrews.ac.uk}

\begin{abstract} 

We show that a wide-field {\it Kepler}-like satellite in Solar orbit
could obtain microlens parallaxes for several thousand events per
year that are identified from the ground, yielding masses and distances for
several dozen planetary events.  This is roughly an order of
magnitude larger than previously-considered narrow-angle designs.
Such a satellite would, in addition, roughly double the number of planet
detections (and mass/distance determinations).  It would also yield
a trove of brown-dwarf binaries with masses, and distances and (frequently)
full orbits, enable new probes of the stellar mass function, and identify
isolated black-hole candidates.  We show that the actual
{\it Kepler} satellite, even with degraded pointing, can demonstrate
these capabilities and make substantial initial inroads into the science
potential.  We discuss several ``Deltas'' to the {\it Kepler} satellite
aimed at optimizing microlens parallax capabilities.  Most of these
would reduce costs.  The wide-angle
approach advocated here has only recently become superior to the
old narrow-angle approach, due to the much larger number of ground-based 
microlensing events now being discovered.

\end{abstract}


\section{{Introduction}
\label{sec:intro}}

Microlens parallaxes have become a crucial focus of microlensing
experiments.  For the overwhelming majority of microlensing events,
one does not know the lens mass, distance, or velocity separately,
but only the Einstein timescale $t_\e$, which is a combination of these
\begin{equation}
t_\e = {\theta_\e\over \mu_\geo}; \quad
\theta_\e^2 = \kappa M\pi_\rel; \quad
\kappa \equiv {4 G\over c^2\,\au}\simeq 8.1\,{{\rm mas}\over M_\odot}.
\label{eqn:tedef}
\end{equation}
Here $M$ is the lens mass, $\theta_\e$ is the Einstein radius,
$\pi_\rel$ is the lens-source relative parallax, and $\mu_\geo$
is the lens-source relative proper motion in the geocentric frame at
the peak of the event.

Microlens parallaxes have become a focus because of the increasing
number of microlens planet detections.  If the planet can be detected
well enough to be characterized, it is almost always because the
source star passed over or very near a ``caustic'' structure induced
by the planet.  This enables one to measure the size of the source radius
relative to the Einstein radius: $\rho = \theta_*/\theta_\e$.  Since
$\theta_*$ itself is routinely measurable from the position of the
source on an instrumental color-magnitude diagram \citep{yoo04},
this means that $\theta_\e$ is measured in almost all planetary events.

Hence, if the microlens parallax
\begin{equation}
\pi_\e = {\pi_\rel\over\theta_\e}; \quad
M = {\theta_\e\over \kappa \pi_\e}
\label{eqn:piedef2}
\end{equation}
could be routinely measured, it would yield masses and distances for
essentially all microlens planets.  Such mass measurements would
yield detailed features in the planet mass function that are
currently washed out by the $M\sin i$ degeneracy for Doppler (RV)
detected planets, and by severe selection for current
transit surveys.  Because microlensing planets are detected from
the Galactic bulge almost to the solar neighborhood, the distance
measurements would yield the Galactic distribution of planets,
and would in particular test for relative planet frequency in
the very different environments of the Galactic disk and bulge.

Unfortunately, microlens parallax measurements today are anything but
``routine''.  As specified in Equation~(\ref{eqn:piedef2}), $\pi_\e$
quantifies the size of the trigonometric lens-source parallax 
oscillations (due to Earth orbit) compared to the Einstein radius.
In principle the photometric effect of these oscillations should just
scale with $\pi_\e$.  However, because $t_\e\ll \rm yr$ for most
microlensing events, the lens-source path is mainly just displaced
from what would be seen from the Sun, rather than oscillating about it.
Hence, $\pi_\e$ measurements are both rare and very heavily biased toward
long events.

However, a quarter century before \citet{gould92} proposed measuring
$\pi_\e$ from such annual oscillations, \citet{refsdal66} had already
pointed to a way out: put a satellite in solar orbit to monitor the
microlensing event {\it simultaneously} with Earth-bound observations.  There
are a number of complications introduced by this geometry 
\citep{refsdal66,gould94,gould95}, which we review below, but these
have been systematically addressed in subsequent work and are now
generally regarded as tractable.

The key obstacle to launching such a satellite was that the potential
scientific return was too modest to justify its substantial cost.
At the time that microlens parallax satellites were first proposed 
circa 1996, relatively few microlensing events were being discovered 
and almost none had $\theta_\e$ measurements.  Hence, $\pi_\e$
measurements would not yield masses or distances
but only somewhat lower statistical errors on Bayesian mass estimates.  
In addition, no planets were being discovered, so that any masses
that were measured would be overwhelmingly of luminous stars, whose
mass function is more easily probed by photometric surveys.

In the meantime, the rate of microlensing event detections has grown
by almost 2 orders of magnitude and continues to grow.  Yet the
basic concept of a parallax satellite equipped with a narrow-angle
camera that would cycle through a list of ongoing events has remained
the same.

Here we argue that a wide-field camera in solar orbit, similar to {\it Kepler},
is a far better match to the requirements of microlensing parallaxes.
Its multiplexing capabilities more than compensate for the inevitable
increase in ``sky'' background due to the larger 
pixels and point spread
function (PSF) that are required to continuously monitor tens of
square degrees.  In addition, by carrying out continuous monitoring
from a position that is well separated in the Einstein ring, the
satellite would essentially double the number of planets discovered.
Finally, such a parallax satellite would be uniquely capable of
many other investigations, such as studies of brown dwarf binaries.

\section{{Nature of Microlensing Parallax}
\label{sec:nature}}

\begin{figure}
\includegraphics[scale=0.72]{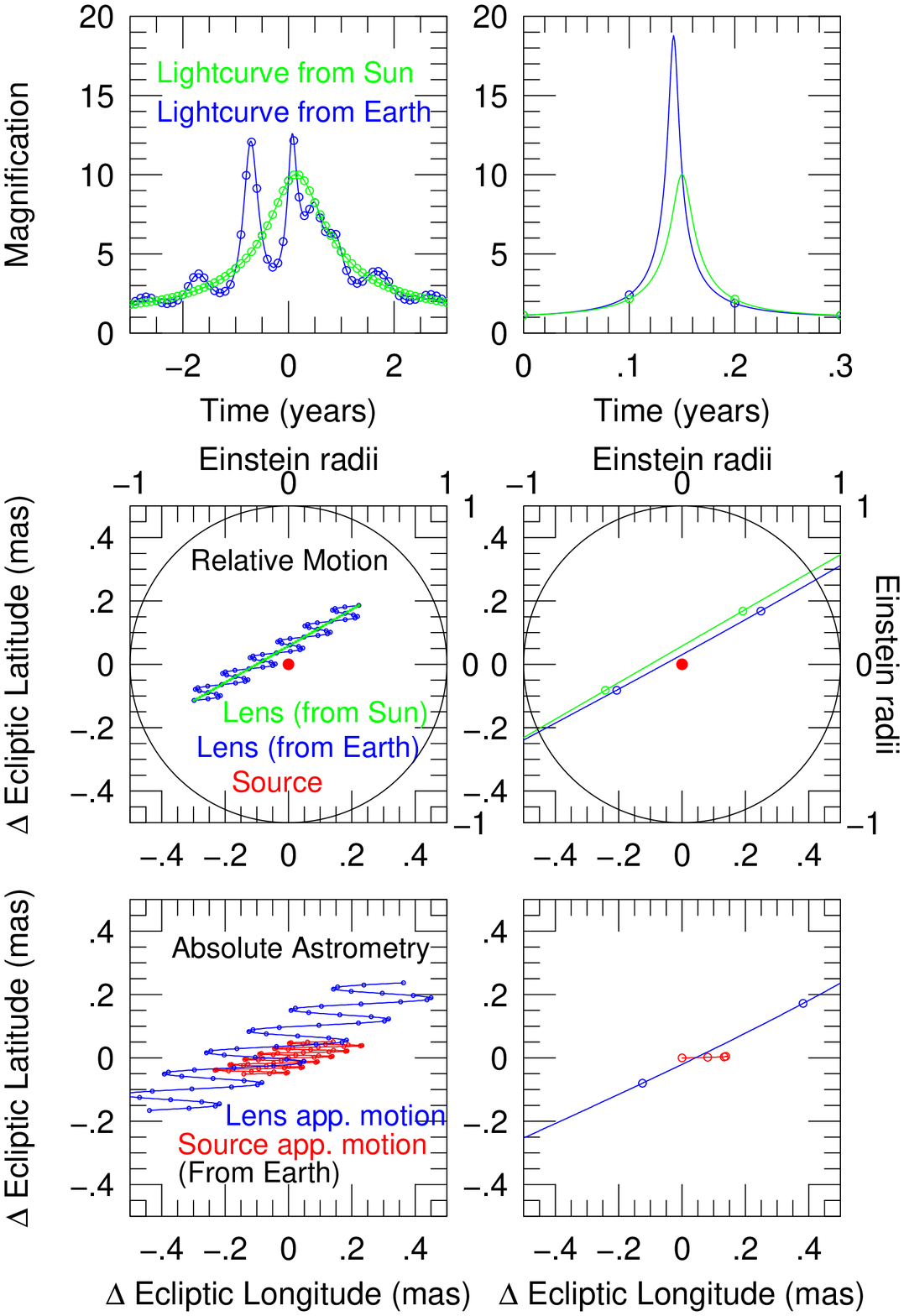}
\caption{\label{fig:traj}
Parallax effect for illustrative (left) and realistic (right) microlensing
events.  Bottom: absolute trigonometric parallax and proper motion (ppm).  
Middle: relative trigonometric (lower/left labels) and microlensing
(upper/right labels) ppm. Top: resulting lightcurves from Earth (blue) and
Sun (green).
}
\end{figure}

Figure \ref{fig:traj} illustrates the relation of microlensing
to trigonometric parallax.  The bottom panels show the absolute astrometry
due to
trigonometric parallax and proper motion (ppm) of the source and lens
separately.  The middle panels show relative ppm (both
trigonometric and microlensing, but with different scaling).  The top
panels show the resulting lightcurves as seen from both the Sun and Earth.  
Both columns have 
$\pi_\rel=60\,\muas$ and $M=0.5\,M_\odot$ (typical values for events
with disk lenses and bulge sources), corresponding to
$\theta_\e=0.5\,{\rm mas}$ and $\pi_\e=0.12$.  The right column
has a heliocentric proper motion $\mu_\hel=5\,\masyr$ (also typical),
while the left column has a highly atypical $\mu_\hel=0.1\,\masyr$,
which only occurs in about $10^{-5}$ of all events.

The left column illustrates the basic principle, but the right column
illustrates the basic problem: in typical events the lens does not
oscillate about its position as seen from the Sun, but is primarily
displaced from it (easily seen in plot but unobservable from Earth).  
The observable effect is then a slight distortion 
(imperceptible here) to the light curve, rather than oscillations.
The figure makes clear why microlensing parallax is a vector
\begin{equation}
\bpi_{\e,\geo} = \pi_\e{\bmu_\geo\over \mu_\geo};
\quad
\bpi_{\e,\hel} = \pi_\e{\bmu_\hel\over \mu_\hel},
\label{eqn:vecpie}
\end{equation}
for which the proper motion provides the direction,
whereas trigonometric parallax is a scalar.  That is, in microlensing
the direction is known only if the parallax is measured.  It also
makes clear why the ``microlensing proper motion'' ($\mu_\e\equiv t_\e^{-1}$)
is a scalar and is measured in the geocentric frame: $\mu_\e=\mu_\geo/\theta_\e$.
That is, in the absence of parallax information (the usual case), there
is only information about the geocentric rate of passage through
the Einstein ring.

Finally, this figure makes clear why one wants to go to solar orbit: by
simultaneously observing the event from two positions indicated by the
blue and green lines, one will see each event from a different perspective
even though one is still deprived of seeing wiggles in the lightcurve.
Although the green line in the figure represents the position of the Sun,
positions in solar orbit have similar separations from Earth.

Figure~\ref{fig:geom} shows an idealized view of the perspectives
seen from Earth and a satellite with zero relative velocity.  From the
lightcurves (bottom panel) one can derive event parameters $(t_0,|u_0|,t_\e)$
for each separately, i.e., time of peak, impact parameter, Einstein timescale.
These yield the vector parallax (separation of blue and red symbols in
upper panel),
\begin{equation}
\bpi_\e = {\au\over D_{\rm sat}}(\Delta\tau,\Delta\beta)
\qquad
\Delta\tau\equiv {t_{0,\rm sat}-t_{0,\oplus}\over t_\e};
\qquad \Delta\beta \equiv u_{0,\rm sat}-u_{0,\oplus}.
\label{eqn:vecpar}
\end{equation}
However, since $u_0$ is a signed quantity while the (easy) observable 
$|u_0|$ is not, $\Delta\beta$ is subject to a four-fold degeneracy depending
on whether the lens passes on the same or opposite side of the source
$(\Delta\beta_\mp)$ from Earth and satellite, and whether it passes on
the right or left $(\pm\Delta\beta)$ as seen from Earth (top panel).
These degeneracies have been extensively
analyzed in the literature \citep{refsdal66,gould94,gould95,gould99,gaudi97,
dong07}.  We discuss in Section~\ref{sec:pathfinder}
how they would be broken for the present case.

\section{{(MP)$^3$: Multi-Plexing for Mass-Production of
Microlens Parallaxes}
\label{sec:mp3}}

Early bulge lensing surveys found several dozen events per season
over several tens of square degrees.  Hence, of order a dozen were
significantly magnified at any given time.  In these conditions, a
narrow-angle camera that cycles through ongoing events was the
obvious choice.  At present, over 2000 events are being discovered
per year, and this is likely to roughly double over the next several
years.  Keeping up with so many events would stress the narrow-angle
approach beyond its limits.

Any practical wide-angle approach will be pixel-limited due to
power, communication, weight, and cost issues.  Because typical events
are $I\sim 17$ while the ``sky'' (actually mean stellar light) is
$I\sim 17.5\,\rm mag\,arcsec^{-2}$, this implies that essentially
all observations will be background limited.  The signal-to-noise ratio
(S/N) for a fixed time interval and fixed pixel number then scales as
\begin{equation}
({\rm S/N})^2 \propto D^2{f\over\Omega_{\rm cam}},
\label{eqn:sn}
\end{equation}
where $D$ is the mirror diameter, $\Omega_{\rm cam}$ is the camera field of
view, and $f$ is the fraction of time spent observing a given field.
Even if there were no overhead for pointing, $f=\Omega_{\rm cam}/\Omega_{\rm tot}$
where $\Omega_{\rm tot}$ is the total field being observed, which implies
S/N$\propto D\,\Omega_{\rm tot}^{-1/2}$.  Since in practice there are such
overheads, one is driven immediately to a {\it Kepler}-like design,
with pixels large enough to cover $\Omega_{\rm tot}$ in a single pointing.
The mirror size can then be adjusted to achieve the desired S/N, keeping
in mind that smaller mirrors put less demands on the optics and weight.

In practice, the great majority of microlensing events are found in a
$\sim 25\,{\rm deg}^2$ area toward the south Galactic bulge.  We will
show below that the {\it Kepler} satellite itself would give quite adequate
performance.  Since the required $\Omega_{\rm cam}$ is about 4 times smaller
compared to {\it Kepler}, the same S/N could be maintained with half the
mirror diameter.  Hence, $D\,\Omega^{1/2}$ (the parameter that basically
quantifies optical design challenges) would be 4 times smaller.
We discuss additional ``Deltas'' with respect to {\it Kepler} in
Section~\ref{sec:discuss}.

\section{{{\it Kepler}: Pathfinder for (MP)$^3$}
\label{sec:pathfinder}}

As already indicated in Section~\ref{sec:mp3}, an optimally designed
microlens parallax satellite would not look exactly like {\it Kepler},
and in particular would be substantially cheaper.  Nevertheless,
{\it Kepler} does exist and is looking for other missions, now that
its pointing stability is no longer adequate for its original mission.
We show here that {\it Kepler}, even with degraded stability, could
make substantial inroads into microlens parallax science.  Equally
important, by carrying out microlensing observations now, {\it Kepler}
would serve as a pathfinder for a future dedicated (MP)$^3$ mission,
in particular providing invaluable aid to design and trade studies.
At the same time, by illustrating how well {\it Kepler} is likely
to function even in its degraded state, we hope to make clear
the potential for additional cost reductions relative to {\it Kepler}.

\begin{figure}
\includegraphics[scale=0.72]{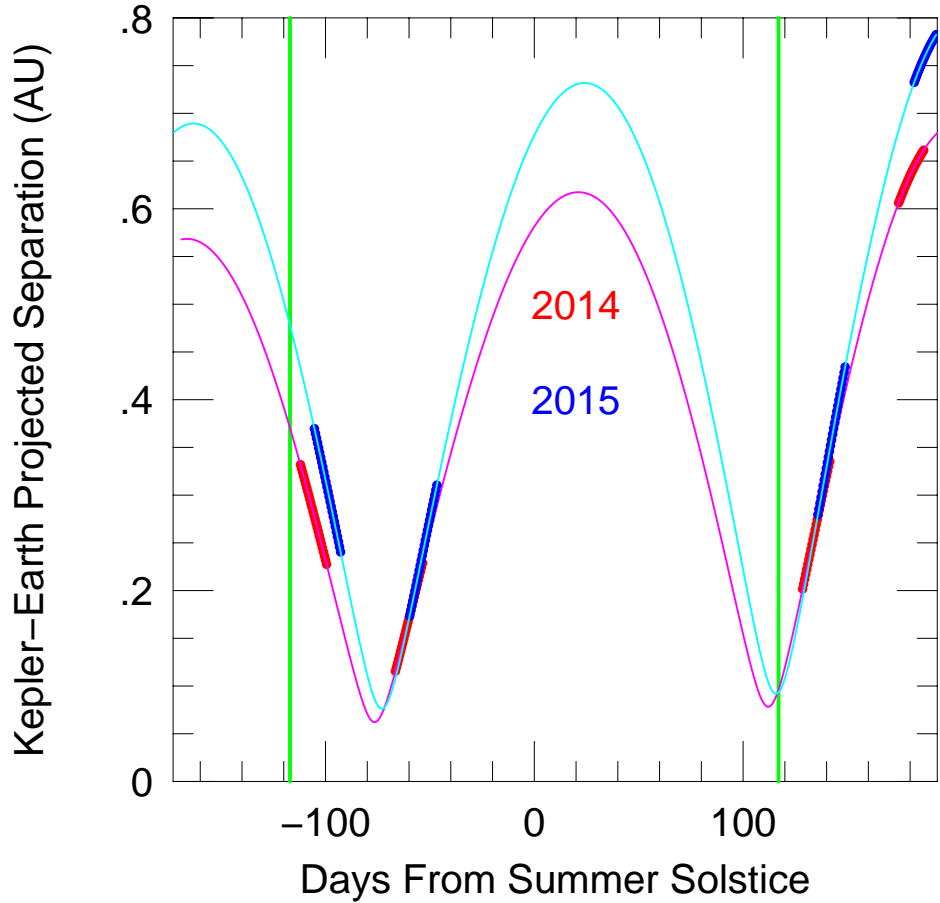}
\caption{{\it Kepler}-Earth projected separation in 2014 
(magenta line)
and 2015 (cyan line) for Galactic Bulge observations.  Red and Blue
points show the times when the boresight can be pointed $\pm 45^\circ$
and $\pm 90^\circ$ from the Sun with the field center still contained
within the field of view.  Green vertical lines show the approximate
boundaries of the microlensing season from Earth.
}
\label{fig:orbit}
\end{figure}

For this purpose, we make the conservative assumption that, since
the bulge fields are near the ecliptic, the boresight must be pointed
exactly at one of four angles relative to the Sun: $\pm 45^\circ$ or 
$\pm 90^\circ$ as originally outlined in the call for white 
papers\footnote{http://keplergo.arc.nasa.gov/News.shtml\#TwoWheelWhitePaper
 . We discuss the implications of rapidly evolving ideas on {\it Kepler} capabilities at the end of this section.}.
Because of the finite size of the {\it Kepler} field of view, this actually
means that the center of the bulge field could be observed continuously 
for about 14 days at each of 4 epochs per year.  From 
Figure~\ref{fig:orbit}, it is clear
that for two of these epochs the bulge is not observable from Earth, so
we consider only the two on the left (except to allow for baseline
observations).  Most targets will be relatively
near the center of the {\it Kepler} field, where the PSF has a FWHM of
$\theta_\fwhm =3.1^{\prime\prime}$.  We adopt 5 minute exposures to minimize
trailing $(0.9^{\prime\prime}/\rm min)$.  Hence, 
the images would be trailed by $\epsilon =4.5^{\prime\prime}$.  Thus, the
effective background area in the oversampled limit is 
$(\pi/0.70\ln 4)\,\theta_\fwhm^2=30\,{\rm arcsec^2}$
\citep{gouldyee13}.  Since a pixel
is 2 times smaller than this number, the oversampled limit is too 
generous: we adopt $40\,{\rm arcsec^2}$.  We assume 3700 photons per second
at $I=15$ (using the {\it Kepler} response function and the conservative
example of a $T=5800$K star with $E(V-I)=2$).

Figure~\ref{fig:lcs} simulates two events, each with source flux $I=18.4$ 
(relatively faint compared to typical targets).  The above parameters lead to 
errors of 0.05 mag per exposure or 0.003 mag per day.  The left-hand
panel shows a typical disk-lens/bulge-source 
event (relatively large $\pi_\e$) and the
right-hand panel shows a typical bulge-bulge event.  The residual
panels show how well the four-fold degeneracy can be broken from the
combination of ground and {\it Kepler} observations.  As can be seen
from Figure~\ref{fig:orbit}, the Earth-satellite separation does not
actually stay fixed (as was assumed in the schematic Figure~\ref{fig:geom}).
This is responsible for most of the degeneracy breaking.  Accelerated motion
of Earth is another such effect.  

Given these conservative 
estimates of {\it Kepler}'s capabilities, at least 300 targets
could be simultaneously monitored.  For a fully functioning {\it Kepler},
this would increase to about 30,000, far above the number of events that
are being discovered.  Recently it has been suggested that {\it Kepler} can
achieve high pointing stability if it stares {\it exactly} at the ecliptic,
and that this can be sustained for 90-day continuous intervals.  Since 
the microlensing fields are centered about $5.5^\circ$ from the ecliptic,
this would allow roughly 60\% of the microlensing events that are discovered
over almost 3 months to be monitored, i.e., roughly 3 times more than was
outlined above.

\section{Discussion
\label{sec:discuss}}

\begin{figure}
\includegraphics[scale=0.72]{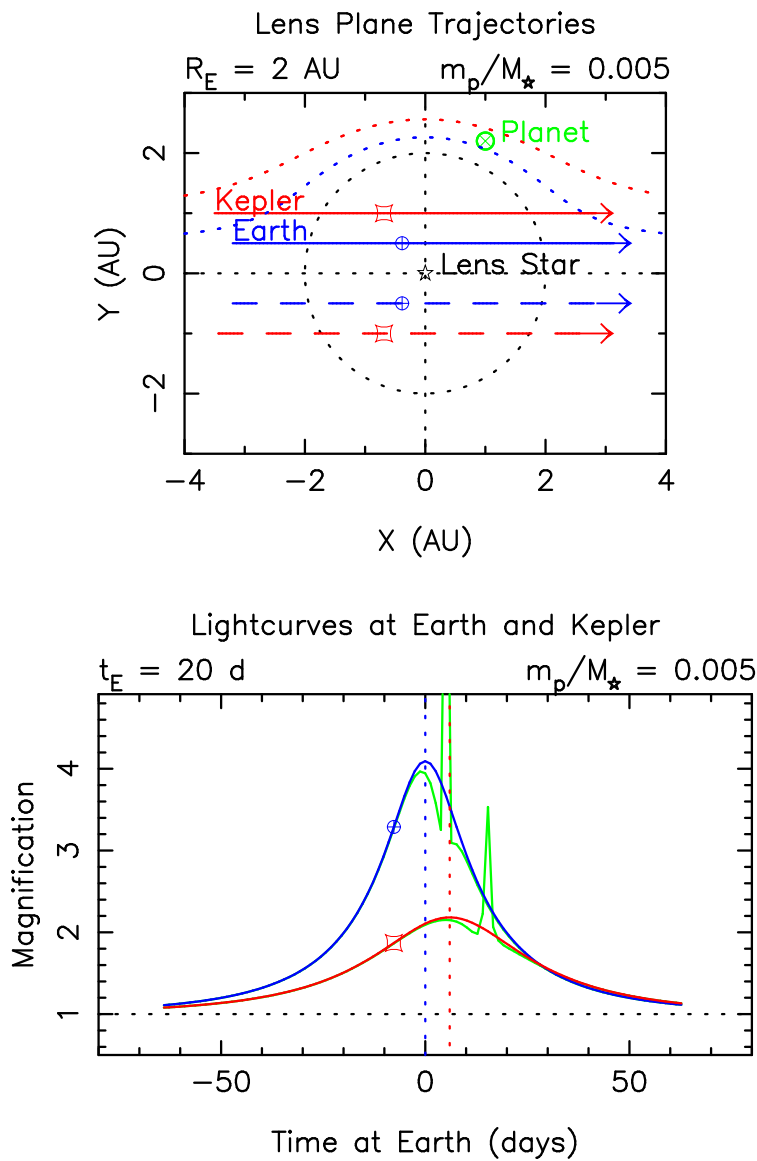}
\caption{\label{fig:geom}
Illustration of four-fold degeneracy derived from comparison of {\it Kepler}
and ground based lightcurves.  Upper panel shows two possible trajectories
of the source relative to the lens for each of {\it Kepler} (red) and Earth 
(blue) observatories.  Each set would give rise to the same point-lens 
lightcurve in the lower panel (same colors), leading to an ambiguity
in the Earth-{\it Kepler} separation (distance between red circle and blue
square) relative to the Einstein ring.  In this particular case, the planet
causes deviations to both lightcurves (green), thus proving that the
trajectories are on the same side of the Einstein ring.  More generally,
the planet would appear in only one curve, leaving the ambiguity open.
In this case, it would be resolved by more subtle differences in the
Einstein timescale.  See Figure~\ref{fig:lcs}, below.
}
\end{figure}

\begin{figure}
 \centering
\begin{tabular}{cc}
 \includegraphics[width=8.5cm, angle=0]{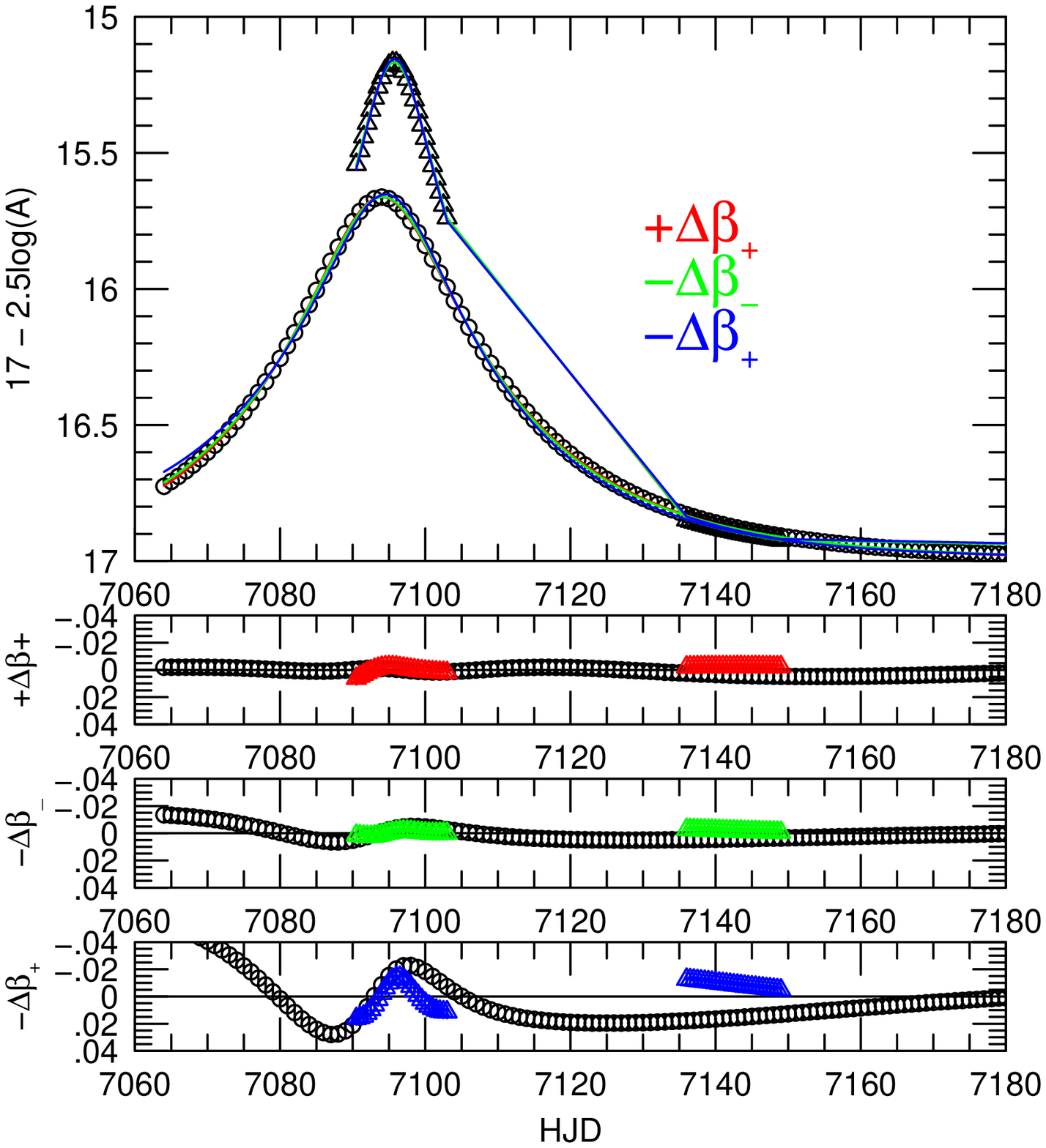}
&
 \includegraphics[width=8.5cm, angle=0]{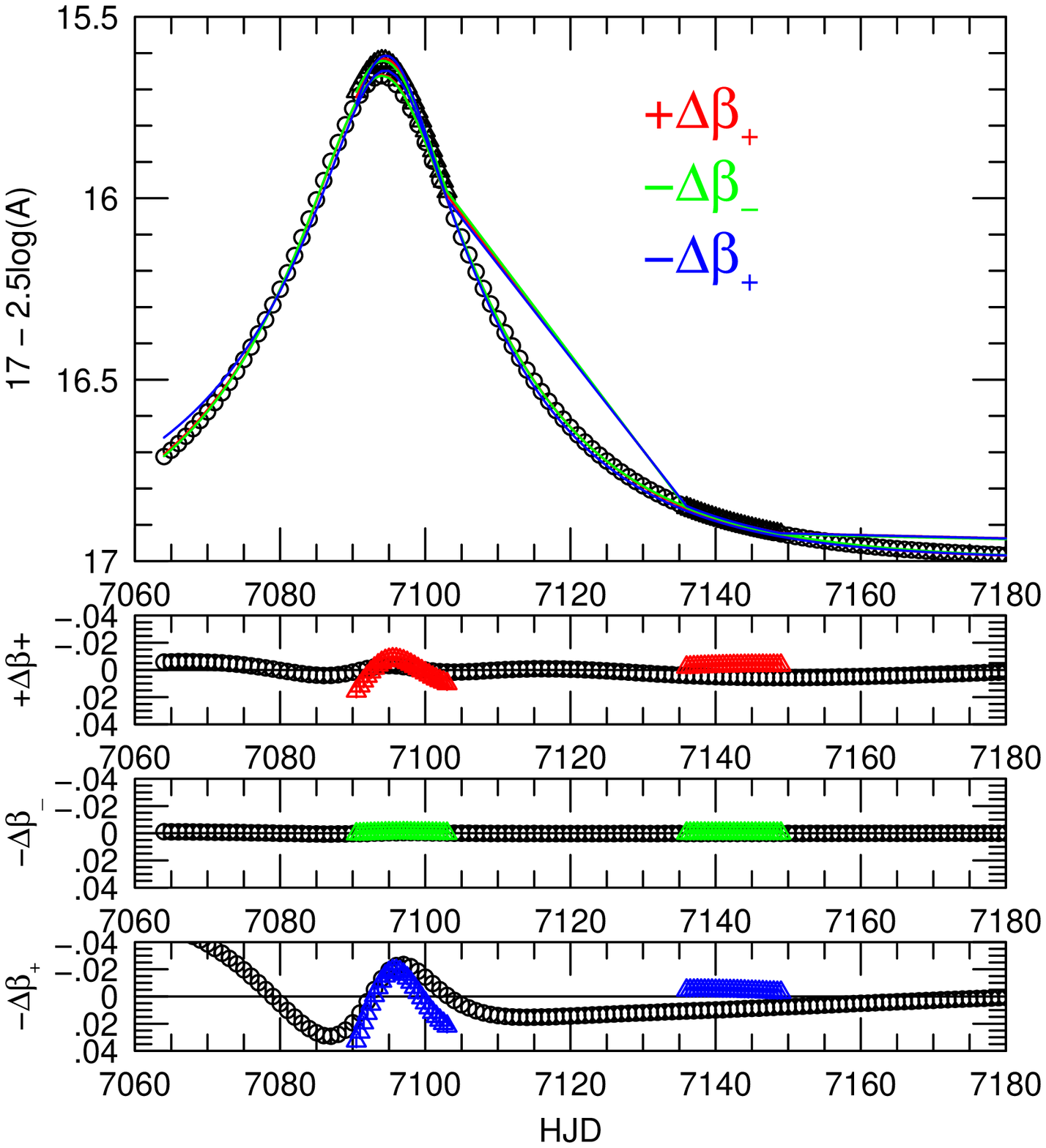}
\end{tabular}
\caption{Parallax measurement and degeneracy-breaking for a typical
Disk-lens event $(\pi_{\e,N},\pi_{\e,E})=(0.3,0.3)$ (left) and Bulge-lens
event $(\pi_{\e,N},\pi_{\e,E})=(0.03,0.03)$ (right). Top panels: 
{\it Kepler} measurements
(triangles) over two 14-day windows and Earth measurements (circles)
over much longer timescale
both shown on arbitrary magnitude scale.  Three different models are shown
for each, indicated at right in different colors.
(In this panel the model curves overlap and can barely be distinguished).
The true model is $+\Delta\beta_-$, i.e., lens passes the source on its
right as seen from Earth (+), and Earth/{\it Kepler} both see the lens passing
the source on the same side (-).  
The error bars are 0.005 mag per day from the ground and 0.003 mag per
day from {\it Kepler}.  Lower Panels: residuals
for each case for Earth (black) and {\it Kepler} (colored). 
{\bf Disk (left):} 
$+\Delta\beta_+$ would have $\pi_\e\sim 1.45$ (factor 3.4 too large) but ruled
out by $\Delta\chi^2=71$;
$-\Delta\beta_-$ would have $\pi_\e\sim 0.44$ (just 5\% too large) and is 
ruled out by
($\Delta\chi^2=124$).
$-\Delta\beta_+$ would have $\pi_\e\sim 1.13$ (factor 2.7 too large) and ruled
out by $\Delta\chi^2=2084$.  
{\bf Bulge (right):} 
$+\Delta\beta_+$ would have $\pi_\e\sim 1.6$ (factor 38 too large) but ruled
out by $\Delta\chi^2=147$;
$-\Delta\beta_-$ would have $\pi_\e\sim 0.048$.  This is permitted
($\Delta\chi^2=1$) but it is just 14\% too large;
$-\Delta\beta_+$ would have $\pi_\e\sim 1.13$ (factor 26 too large) and ruled
out by $\Delta\chi^2=1729$. 
}
\label{fig:lcs}
\end{figure}

Roughly a dozen planets are discovered per year, so even the two
14-day epoch experiment illustrated in Figure~\ref{fig:lcs} would
be expected to measure parallaxes (hence masses) for about two of them.
Moreover, by the same estimate, continuous coverage of all events
(or at least all bright enough to yield planet detections) would
independently detect planets.  In rare cases 
(illustrated in Figure~\ref{fig:geom}), these would be the same planets, but
most often they would be different.  Hence, that experiment would
yield about 4 planet masses.  In the more favorable 
90-day ecliptic-observation scenario,
about 12 planet masses would be measured
(half each discovered from ground and space).  For continuous coverage
over the whole season (not possible with {\it Kepler}) the numbers
would rise to 12 and 24.  Since the number of planet detections
is rising, these numbers would grow as well.

As already pointed out, an optimized (MP)$^3$ mission could have
a mirror with half the {\it Kepler} diameter (or 1/4 of the pixels)
and still do as well as {\it Kepler}.  Taking account of the fact
that (MP)$^3$ would operate continuously (compared to the more limited
coverage illustrated
in Figure~\ref{fig:lcs}), it would permit substantial further reductions.
On the other hand such a mission would need additional solar panels
to permit observations in opposition and would also need a somewhat
stronger ``push'' into Earth-trailing orbit (say $1\,\kms$ to
get 1 AU from Earth in 1.6 years), plus a comparable decelerating thrust
so it remained near that distance from Earth.

(MP)$^3$ would have many science applications in addition to microlensing
planets.  Essentially all binary lenses would yield mass/distance
measurements, allowing nearly perfect identification of systems in
which one or both components are dim or dark (i.e., brown dwarfs, neutron
stars, or black holes).  In particular, double-brown-dwarf systems are
only so-identified in unusually favorable circumstances in present
microlensing searches \citep{choi13} and, particularly at the low-mass
end, are difficult or impossible to study by other techniques.  Because
these have much larger caustics than planetary systems, they would
be observed from both Earth and the satellite (at different times), leading
most often to complete orbital solutions, which again is extremely rare
in current Earth-bound surveys \citep{shin11,shin12}.  

Finally, 
candidate isolated black holes would be routinely identified from their
relatively small parallaxes $\pi_\e = \sqrt{\pi_\rel/\kappa M}$ and
long timescales $t_\e = \theta_\e/\mu_\geo = \sqrt{\kappa M\pi_\rel}/\mu_\geo$.
These could subsequently be distinguished from ordinary stars (normal
$\theta_\e$, exceptionally low $\mu_\geo$) because their large $\theta_\e$ would
give rise measurable astrometric effects in high-resolution followup
observations \citep{my95,hnp95,walker95}.

\acknowledgments

We thank Scott Gaudi and Jennifer Yee for stimulating discussions.
Work by AG was supported by NSF grant AST 1103471 
and NASA grant NNX12AB99G.  
KH is supported by a Royal Society Leverhulme Trust Research Fellowship
and by grant NPRP-09-476-1-78 from the Qatar National Research Fund (a
member of Qatar Foundation).

\end{document}